\RequirePackage{lineno}
\documentclass[prl,twocolumn,showpacs,superscriptaddress]{revtex4}

\usepackage{graphicx}
\usepackage{amsmath}
\usepackage{amssymb}
\usepackage{epsfig}
\usepackage{epstopdf}

\renewcommand{\v}[1]{{\bf #1}}

\newcommand{\xh}{{\hat{x}}}
\newcommand{\yh}{{\hat{y}}}

\def\eqa{\begin{eqnarray}}
\def\eea{\end{eqnarray}}
\newcommand{\eq}{\begin{equation}}
\newcommand{\ee}{\end{equation}}
\newcommand{\nn}{\nonumber\\}

\renewcommand{\Im}{{\rm Im}}
\renewcommand{\Re}{{\rm Re}}

\newcommand{\ua}{\uparrow}
\newcommand{\da}{\downarrow}

\newcommand{\al}{\alpha}
\newcommand{\bt}{\beta}
\newcommand{\del}{\delta}
\newcommand{\Del}{\Delta}
\newcommand{\eps}{\epsilon}

\newcommand{\ga}{\gamma}
\newcommand{\Ga}{\Gamma}

\newcommand{\La}{\Lambda}

\newcommand{\si}{\sigma}

\usepackage{color}
\begin{document}
%\linenumbers
%\leftlinenumbers*
%\rightlinenumbers*

\title{Theory of the Evolution of Superconductivity in Sr$_2$RuO$_4$ under Anisotropic Strain}

\author{Yuan-Chun Liu}
\affiliation{National Laboratory of Solid State Microstructures $\&$ School of Physics, Nanjing
University, Nanjing, 210093, China}

\author{Fu-Chun Zhang\footnote{fuchun@hku.hk}}
\affiliation{Department of Physics, Zhejiang University, Hangzhou 310027, China}
%\affiliation{Collaborative Innovation Center of Advanced Microstructures, Nanjing 210093, China}

\author{Thomas Maurice Rice\footnote{rice@phys.ethz.ch}}
\affiliation{Institute for Theoretical Physics, ETH H\"onggerberg, CH-8093  Z\"urich, Switzerland}

\author{Qiang-Hua Wang\footnote{qhwang@nju.edu.cn}}
\affiliation{National Laboratory of Solid State Microstructures $\&$ School of Physics, Nanjing
University, Nanjing, 210093, China}
\affiliation{Collaborative Innovation Center of Advanced Microstructures, Nanjing University, Nanjing 210093, China}
%\date{\today}

\pacs{74.20.-z, 74.20.Rp, 71.27.+a}
%
%75.30.Fv  Spin-density waves
%74.20.Rp  Pairing symmetries (other than s-wave)
%74.20.-z  Theories and models of superconducting state
%71.27.+a  Strongly correlated electron systems; heavy fermions
%64.60.ae  Renormalization-group theory

\maketitle

{\em Abstract}: {\bf Sr$_2$RuO$_4$ is a leading candidate for chiral $p$-wave superconductivity. The detailed mechanism of superconductivity in this material is still the subject of intense investigations. Since superconductivity is sensitive to the topology of the Fermi surface (the contour of zero-energy quasi-particle excitations in the momentum space in the normal state), changing this topology can provide a strong test of theory. Recent experiments tuned the Fermi surface topology efficiently by applying planar anisotropic strain. Using functional renormalization group theory, we study the superconductivity and competing orders in Sr$_2$RuO$_4$ under strain. We find a rapid initial increase in the superconducting transition temperature $T_c$, which can be associated with the evolution of the Fermi surface toward a Lifshitz reconstruction under increasing strain. Before the Lifshitz reconstruction is reached, however, the system switches from the superconducting state to a spin density wave state. The theory agrees well with recent strain experiments showing an enhancement of $T_c$ followed by an intriguing sudden drop.}\\

{\em Introduction}: Chiral $p+ip'$-wave superconductor is currently of great research interest because of its topological property that may lead to zero energy Majorana bound states~\cite{ivanov,green}, the building block for topological quantum computing~\cite{nayak}. However, this type of intrinsic superconductivity is rare. The leading candidate to date is Sr$_2$RuO$_4$ \cite{mackenzie, bergemann, Maeno2012, Kallin2012} discovered more than twenty years ago~\cite{maeno}. In agreement with the initial theoretical proposals ~\cite{ricesigrist,baskaran}, various experiments provide evidence for odd parity Cooper pairs~\cite{liu}, with total spin equal to one~\cite{NMR}, and  chiral time-reversal breaking symmetry~\cite{muSR,Kerr}. The Fermi surface of Sr$_2$RuO$_4$  has two distinct components \cite{mackenzie,bergemann}, two approximately 1-dimensional (1D) $\alpha$ and $\beta$ bands and a single 2D $\gamma$ band. There has been a continuing discussion for some years as to which components determine the value of $T_c$ \cite{others,Kivelsona,Kivelsonb,Lederer}. Early specific heat measurements and calculations pointed to different onset temperatures for pairing in the three relevant bands, suggesting that only a subset of the bands dominate the superconductivity~\cite{Agterberg97}. Functional renormalization group (FRG) calculations based on the full set of three bands strongly supported the proposal that $T_c$ is determined predominantly by the single 2D $\gamma$ band \cite{sro3orb}, while the other bands are coupled to the $\ga$-band weakly by inter-band Josephson effect \cite{Sigrist}. However the expected edge current in a chiral superconducting state has not been conclusively detected~\cite{kirtley}, which led to the proposal that the $p$-wave pairing might arise from the $\al$ and $\bt$ bands, with exact cancelation of charge topological numbers from the hole and electron pockets in these bands~\cite{Kivelsona,Kivelsonb}.

Recent experiment by Hicks {\it et al} \cite{hicks} and most recently by Steppke {\it et al} \cite{steppke} found that the transition temperature $T_c$ in the $p$-wave superconductor Sr$_2$RuO$_4$ rises dramatically under the application of a planar anisotropic strain, followed by a sudden drop beyond a larger strain. A special feature of the 2D $\gamma$ Fermi surface is the very close approach to the 2D van Hove singularities (vHS) at the X/Y points. As a result a strong effect on $T_c$ can be expected in view of the normal state density of states (DOS), since an anisotropic distortion drives the Fermi surface towards the vHS along one axis ({\it e.g.}, Y), and away from the vHS in the perpendicular direction ({\it e.g.}, X), leading to Lifschitz reconstruction of the Fermi surface. The sensitivity of $T_c$ versus strain thus points to the dominant role of the $\ga$ band. The more intriguing sudden drop of $T_c$ calls for an interpretation beyond the simple DOS argument. Therefore, aside from the interest in the increased $T_c$, the experiment also provides a unique test for theory.\\

\begin{figure}
\includegraphics[width=8cm]{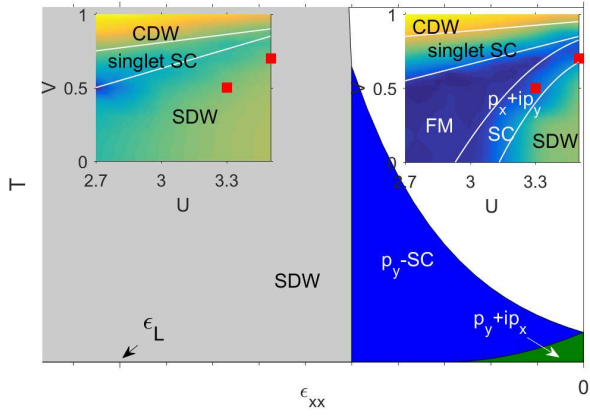}
\caption{Main panel: Schematic phase diagram in the temperature $T$ - strain $\eps_{xx}$ space (with $\eps_{yy}\equiv -\eps_{xx}>0$) under fixed interaction. Insets: the phase diagram in the interaction parameter ($U$ - $V$) space for $\eps_{xx}=\eps_L$ (left inset) and $\eps_{xx}=0$ (right inset). Here $\eps_L$ (arrow) is the strain level for Lifshitz reconstruction. The ordering temperature of the various phases in the insets is higher where the background color is brighter. The white solid lines denote qualitatively the phase boundaries, and the red squares denote the two typical sets of interactions discussed in more details in the text.}
\end{figure}

{\em Results}: Here we apply FRG to study the effect of anisotropic strain in the tetragonal RuO$_2$ planes. Since the earlier calculation based on three bands for the unstrained system indicated the 2D $\ga$ band is active \cite{sro3orb}, and since only the $\ga$ band responses to the strain sensitively, we limit our FRG calculations to this band. Our results are summarized schematically in Fig.1 (main panel). We find a clear initial increase of $T_c$ under the strain, consistent with the observation of Steppke {\it et al}~\cite{steppke}. More importantly, our theory explains the sudden drop of $T_c$ by a transition into a spin density wave (SDW) state before the Lifshitz transition (at the strain level $\eps_L$) is reached, a prediction that can be tested by nuclear magnetic resonance (NMR) experiments. We also point out a second phase transition from a time-reversal invariant (T-invariant) $p$-wave pairing ($p_y$ for $\eps_{xx}<0$) at higher temperatures to a time-reversal symmetry breaking (T-breaking) $p_y+ip_x$-type pairing at lower temperatures in strained Sr$_2$RuO$_4$. We find singlet pairing is unlikely.\\

{\em Discussion}: We consider a single band extended Hubbard model in a 2D square lattice to describe the relevant $\gamma$ band arising from $d_{xy}$ orbital in Sr$_2$RuO$_4$.  The Hamiltonian reads,
\eqa
H = &&-\sum_{i,\si,\v b=\xh,\yh}(t+\eta_\v b\del t)(c_{i\si}^\dag c_{i+\v b,\si}+{\rm h.c.})\nn
&&-t'\sum_{i,\si,\v b=\xh\pm \yh}(c_{i\si}^\dag c_{i+\v b,\si}+{\rm h.c.})-\mu\sum_i n_i \nn
&&+U\sum_i n_{i\ua}n_{i\da}+V\sum_{i,\v b=\xh,\yh}n_i n_{i+\v b}. \label{H}
\eea
Here $c^\dag_{i\si}$ creates an electron at site $i$ with spin $\si=\ua$ or $\da$, $\xh$ ($\yh$) is a unit vector along $x$ ($y$) axis. We assume an anisotropic strain $\eps_{xx}\equiv -\eps_{yy}<0$ for definiteness, which leads to a change of the nearest-neighbor (NN) hopping by $\eta_\v b\del t =\pm \del t$ for $\v b=\xh/\yh$ and with $\del t\propto -\eps_{xx}>0$. This single-parameter modeling of the strain effect should be a good approximation for weak strain.\cite{density} The remaining notation is standard. The first two lines describe the free normal state, and the last line the onsite ($U$) and NN ($V$) Coulomb interactions. We set $t = 0.8$, $t' = 0.35$ and $\mu = 1.3$, henceforth in dimensionless units. The resulting Fermi surface matches that of the experimental $\gamma$ band in Sr$_2$RuO$_4$ in the absence of strain (namely $\del t=0$).

\begin{figure}
	\includegraphics[width=7cm]{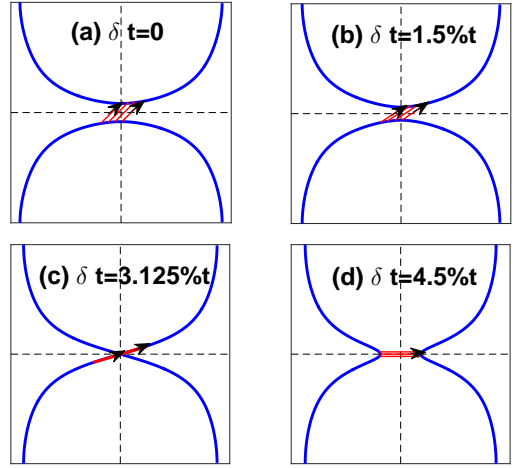}
	\caption{Normal state Fermi surfaces (blue lines) in the Brillouine zone and quasi-nesting vectors (arrows and red bars) for $\del t/t=$ (a) $0$, (b) $1.5\%$, (c) $3.125\%$ and (d) $4.5\%$. Notice the center of the zone (crossing point of the dashed lines) is set at momentum $(0,\pi)$ to highlight the quasi-nesting near the van Hove point.}
\end{figure}

Fig.2 shows the evolution of the Fermi surface as $\del t$ changes. Interestingly, a small $\del t$ (of a few percent of $t$) is enough to change the Fermi surface topology significantly, with a Lifshitz transition at $\del t_L/t=3.125\%$. We set the center of the zone at momentum $(0,\pi)$ in order to illustrate the quasi-nesting (arrows and red bars) near the van Hove point, which leads to enhancement of the susceptibility at the corresponding wavevectors. For example, in Fig.3 we show the $\v q$ dependence of the zero-frequency bare susceptibility $\chi_0(\v q)$ (calculated at $T=0.01$). For $\del t$ below the Lifshitz value $\del t_L$, $\chi_0 (\v q)$ peaks at a small $\v q\sim \v q_1$ along the $(1, 1)$ directions, a value clearly associated with that highlighted in Fig.2. In particular, at zero strain $\v q_1 \sim (\pm 0.18, \pm 0.18)$ as found earlier \cite{smallqa,smallqb,smallqc,smallqd,sro3orb}, values consistent with the neutron scattering data \cite{neutron}, implying such ferromagnetic-like spin fluctuations are further enhanced by interaction. Increasing $\del t$ towards $\del t_L$, leads to $q_{1y}$ shrinking while $q_{1x}$ increases. For $\del t$ well above $\del t_L$ , a large wavevector $\v q_2 \sim (\pm 1, \pm 0.4)$ emerges as the dominant SDW wavevector. In fact $\v q_2$ as a local peak position barely changes with $\del t$, implying that it does not follow from Fermi surface nesting, but from off-shell particle-hole (p-h) excitations at finite energies. These dramatic changes of the Fermi surface and bare susceptibility suggest possible dramatic evolution of superconductivity and competing phases, as addressed below.

\begin{figure}
	\includegraphics[width=8cm]{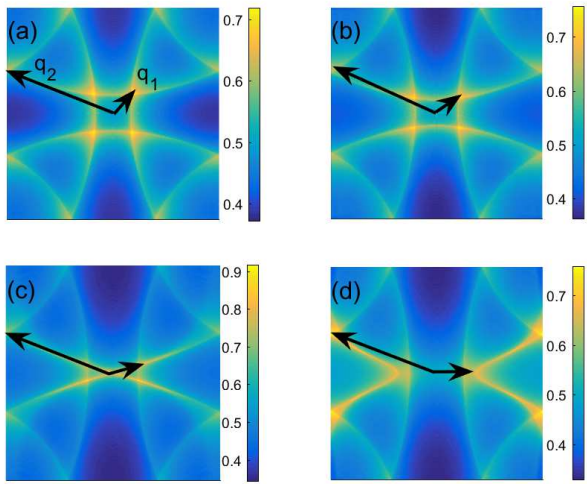}
	\caption{Bare susceptibility $\chi_0(\v q)$ versus the wavevector $\v q$ for $\del t/t=$ (a) $0$, (b) $1.5\%$, (c) $3.125\%$ and (d) $4.5\%$. Here the Brillouine zone is centered at the origin. The arrows indicate strong peaks at a small (large) wavevector $\v q_1$ ($\v q_2$), and $\v q_1$ can be associated with the quasi-nesting vector indicated in Fig.2.}
\end{figure}

We now turn to the discussion of the effect of interactions in driving SC under strain, and the competition among SC, SDW, as well as possible charge density wave (CDW) instabilities. We remark that while a Cooper instability emerges even for an infinitesimal interaction according to the Kohn-Luttinger anomaly, the instability in the particle-hole channels requires a finite interaction in general, except for \emph{accidental} perfect nesting or van Hove singularities on the Fermi surface. As a rough estimate, the Stoner instability requires $\Ga_0\chi_0\geq 1$, where $\Ga_0$ is the appropriate bare interaction for a given particle-hole channel. From $\chi_0$ in Fig.3 we see that the Stoner instability (down to the temperature scale $T=0.01$) requires $\Ga_0\sim 1$. Therefore the competition among various orders is beyond theories in the limit of infinitesimal interactions.\cite{Kivelsona,Kivelsonb,Lederer,steppke} This motivates us to use the singular-mode FRG (SMFRG) \cite{sro3orb,smfrga,smfrgb,smfrgc,husemann}. (See Supplementary Materials for technical details.) Similarly to the more conventional patch-FRG,\cite{patchfrga,patchfrgb,patchfrgc} SMFRG treats competing orders on equal footing and turns out to be reliable up to moderate interactions. Moreover, SMFRG is advantageous for systems near or at the van Hove singularity.\cite{smfrga,smfrgb,smfrgc,husemann} In a nutshell, it extends the usual pseudo-potential for point charges to an effective interaction $\frac{1}{2N}c_{1\si}^\dag c_{2\si'}^\dag\Ga_{1234}c_{3\si'}c_{4\si}$ acting on quasi-particles below an energy scale $\La$, with a general one-particle-irreducible (1PI) vertex function $\Ga_{1234}$. Here $1=\v k_1$ labels the incoming (or outgoing) electrons and $N$ is the number of lattice sites. Momentum conservation and summation over repeated indices are understood. Starting from the bare interaction in Eq.(\ref{H}) at $\La\gg 1$, $\Ga$ flows with decreasing $\La$, picking up 1PI corrections to all orders in the bare interactions. Concurrently, we extract from $\Ga_{1234}$ the effective scattering matrices between fermion bilinears in the SC/SDW/CDW channels, which are subsequently resolved in terms of scattering between eigenmodes at each collective wavevector $\v q$. We take as the representative interaction $V_{\rm X}(\v q)$ the leading attractive eigenvalue (out of many) of the scattering matrix for X = SC/SDW/CDW and at a given $\v q$. In this way, each $V_{\rm X}(\v q)$ is associated with an eigenfunction (i.e., form factor) describing how fermion bilinears are superpositioned within the collective mode. The leading and diverging $V_{\rm X}(\v q)$ at $\La=\La_c$ among all channels implies an emerging order at $T=T_c\sim \La_c$, described by the associated wavevector and form factor. (Order parameter fluctuations may lower $T_c$. In particular, by Mermin-Wagner theorem, long-range orders breaking continuous symmetries are absent at finite temperatures in two dimension, but here they should be understood as stabilized by weak inter-layer couplings in realistic materials.)

\begin{figure}
	\includegraphics[width=8cm]{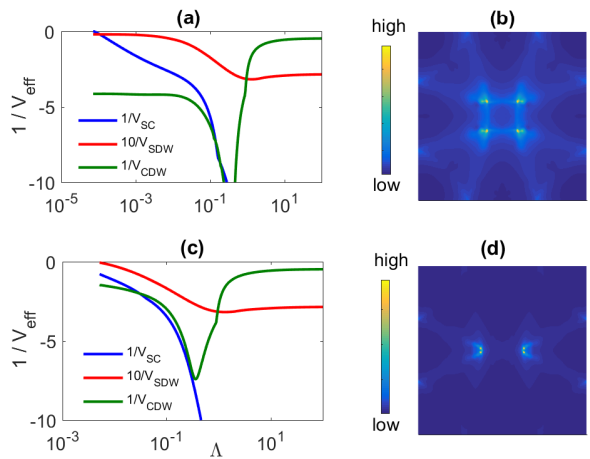}
	\caption{FRG flow versus a decreasing $\La$ for (a) $\del t/t = 0.3\%$ and (c) $\del t/t=3.6\%$. Correspondingly, (b) and (d) show $-V_{{\rm SDW}}(\v q)$ at the late stage of flow.}
\end{figure}

We start the discussion by choosing a set of interactions $(U,V)=(3.5,0.7)$ (see inset in Fig 1) that leads to a triplet chiral $p$-wave superconducting state at zero strain. We define henceforth $V_{\rm X}=\min [V_{\rm X}(\v q)]$ to get an overall view of $V_{\rm X}(\v q)$ for ${\rm X}={\rm SC}/{\rm SDW}/{\rm CDW}$. Fig.4(a) shows the flow of $1/V_{\rm X}$ for $\del t/t=0.3\%$. The SDW channel dominates at higher scales (notice $10/V_{{\rm SDW}}$ is plotted for clarity), is enhanced at intermediate scales, and finally saturates at low scales. Fig.4(b) shows $V_{{\rm SDW}}(\v q)$ in the late stage of the flow. There are strong peaks around small wavevectors, consistent with $\v q_1$ in the bare $\chi_0(\v q)$. We notice, however, that the peak position in $V_{{\rm SDW}}(\v q)$ evolves with $\La$ from large $\v q$ and settles down on small $\v q$. This verifies the earlier argument that $\v q_2$ in $\chi_0(\v q)$ is related to off-shell finite energy p-h excitations. The CDW channel is moderate initially, screened in the intermediate stage, re-enhanced but saturated eventually. The non-monotonic behavior follows from level crossing between different CDW eigenmodes, the details of which are however irrelevant here. The SC channel is repulsive initially (thus out of the field of view), and only becomes attractive in the intermediate stage where the SDW channel is enhanced, a manifestation of channel overlap. Eventually, the SC channel diverges on its own via the Cooper mechanism while the other channels saturate. Fig.4 (c) and (d) are similar plots to (a) and (b) but for $\del t/t=3.6\%$ slightly above the Lifshitz point. The larger $\chi_0(\v q)$ conspires to drive the SDW channel to diverge first. The diverging $V_{\rm SDW}$ corresponds to a diverging renormalized spin susceptibility. We denote such an SDW phase as ${\rm SDW}_1$. Finally for even larger $\del t$ (not shown), we find the SDW channel diverges first again, but now at the large wavevector consistent with $\v q_2$ in $\chi_0(\v q)$, and such a phase is denoted as ${\rm SDW}_2$.

\begin{figure}
	\includegraphics[width=8cm]{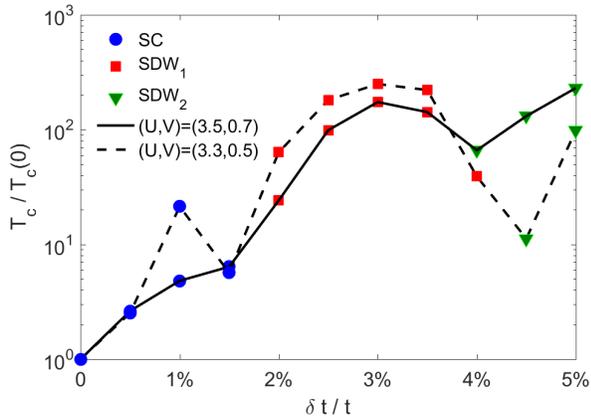}
	\caption{Figure 5: Normalized transition temperature $T_c$ (symbols) versus $\del t$ from FRG calculations based on Eq.(\ref{H}). The lines are guides to the eyes for $(U,V) =(3.5,0.7)$ (solid line) and $(U,V)=(3.3,0.5)$ (dashed line). The symbols indicate the respective phase that would emerge below $T_c$.}
\end{figure}

By systematic calculations, we find qualitatively the same behavior for other values of $(U,V)$, to the extent that {\it in the  unstrained system the $p$-wave SC phase wins} with $T_c> 2\times 10^{-5}$ (or above $0.1$K if we take the bandwidth to be roughly 3eV \cite{xiang,singh}), aiming at a reasonable  comparison with experiment. The range of interaction here is also consistent with the fact that there is a sizable renormalization of the effective mass in Sr$_2$RuO$_4$ \cite{mackenzie,ricesigrist}. As two typical examples, we show in Fig.5 the transition temperature $T_c$ (symbols) of the various phases versus $\del t$ for $(U,V)=(3.5,0.7)$ and $(U,V)=(3.3,0.5)$. In both cases $T_c$ in the SC phase rises rapidly for $\del t<2\%t$, then switches to the SDW$_1$ phase with a peak in $T_c$ around the Lifshitz point $\del t_L$, and finally the SDW$_2$ phase enters for even higher $\del t$. We notice, however, that the SC $T_c$ keeps rising until SDW$_1$ for $(U,V)=(3.5,0.7)$, while a peak is found for $(U,V)=(3.3,0.5)$. We ascribe such difference to the sensitivity of the effect of $\del t$.  The increase of $T_c$ upon strain is in agreement with the experiment of Steppke {\it et al} \cite{steppke}, and our prediction of the transition to SDW phase under strong strain hopefully will be tested in future experiments.

\begin{figure}
	\includegraphics[width=8cm]{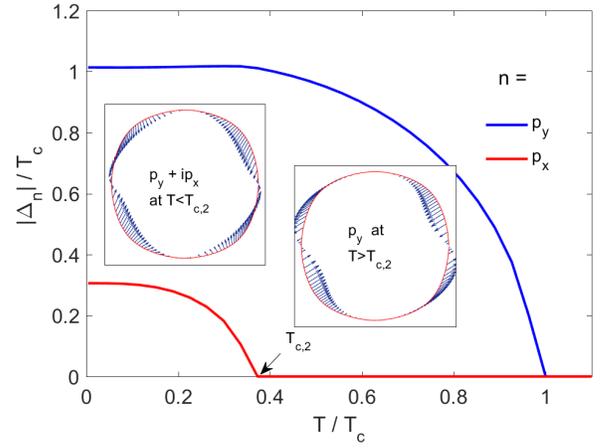}
	\caption{The amplitude of the order parameters, $|\Del_n|$ in the two symmetry channels $n=p_y$ and $n=p_x$,  versus temperature in a FRG-based mean field theory for $\del t=0.3\%t$ and $(U,V)=(3.5,0.7)$. The arrow in the insets show the gap function $\Del_\v k$ as a vector $[\Re \Del_\v k,\Im\Del_\v k]$, for $\v k$ on the Fermi surface. $T_{c,2}$ marks the second SC transition into a T-breaking phase.}
\end{figure}

As the SC channel diverges at $\La=T_c$, the form factor of the leading eigenmode is just the normalized gap function. We find this eigenmode is non-degenerate and respects the $p_y$-wave symmetry, consistent with the $C_{2v}$ symmetry under strain and the fact that $|q_{1y}|<|q_{1x}|$ due to the weakened (enhanced) NN hopping along $y$ ($x$). Note, $p_x$- and $p_y$-wave modes are always degenerate for $\del t=0$ which leads to T-breaking below $T_c$.\cite{sro3orb} A natural question is whether T-breaking still occurs below $T_c$ for $\del t >0$. To answer this question, we extract, at a scale $\La$ slightly above $T_c$, the pairing matrix $V_{p}(\v k,\v k')=\Ga_{\v k,-\v k,-\v k',\v k'}$ acting on electrons for $|\eps_{\v k,\v k'}|\leq \La$, where $\eps_\v k$ is the normal state dispersion. The pairing function $\Del_\v k$ is then determined in close analogy to the Bardeen-Cooper-Schrieffer theory. (See Supplementary Materials.) We call this an FRG-based mean-field theory. Similar approaches have been explored in the literature.\cite{reiss,wangjin} Resolving $\Del_\v k$ into $p_x$- and $p_y$-wave symmetry components, we plot the amplitudes in Fig.6 versus temperature, for $\del t/t=0.3\%$ and $(U,V)=(3.5,0.7)$. Near $T_c$ there is no mixture of symmetries, and this is also clear in the plot of $\Del_\v k$ on the Fermi surface in the right inset. However, symmetry mixing is found deep in the ordered state to be energetically stable below a second transition temperature $T_{c2}$ (arrow in the main panel). In this case, the gap function on the Fermi surface is shown in the left inset, where the clear winding of the phase signals a T-breaking $p_y \pm i \nu p_x$-order, with $0<\nu<1$. The prediction of the second phase transition in our theory has interesting experimental consequences, {\it e.g.} a second jump in the specific heat, and topological states in the vortex core or at the edge in the T-breaking phase. We find $T_{c2}$ decreases rapidly as $\del t$ increases, {\it e.g.} $T_{c2}<0.1 T_c$ already at $\del t=0.5\%t$.

To better understand our result and to compare with the experiment of Ref.\cite{steppke}, in Fig.1 (insets) we have also examined the instability of the system for $\eps_{xx}=0$ (right inset) and $\eps_{xx}=\eps_L$ (left inset) in the $(U, V)$ parameter space. As general tendences, a larger $U$ would favor SDW, while a larger $V$ would favor CDW. At zero strain (right inset), the $p$-wave SC phase is sandwiched by a higher-$T_c$ SDW phase and a lower-$T_c$ ferromagnetic (FM) phase that wins narrowly over the SC phase. (The SC phase may win over FM if the bare interaction is even smaller, but at much lower energy scales.) We remark that if $p$-wave wins at $\del t=0$, the same interaction ({\it e.g.}, the solid red squares) leads to SDW at $\del t=\del t_L$. On the other hand, singlet pairing would appear at both $\del t=0$ and $\del t_L$, {\it provided that $V\sim U/4$}. However, in this case the singlet gap function has lots of nodes, and the normal state would be dominated by CDW fluctuations, which are unlikely given the SDW fluctuations known from neutron scattering \cite{neutron}. In Ref.\cite{steppke}, singlet pairing is proposed to account for the behavior of the upper critical field, which however could also be reconciled in the picture of triplet pairing.\cite{hc2} Moreover, singlet pairing is inconsistent with the earlier strong evidences for triplet pairing.\cite{liu,NMR,muSR,Kerr}

A closer comparison between theory and experiment can be made by referencing our results to the strain level $\eps_L$ for the Lifshitz reconstruction. In our model the transition from SC to SDW$_1$ (and thus a sharp drop of SC $T_c$) occurs between $\eps_{xx}=0$ and $\eps_{xx}=\eps_L$. In the experiments, the sharp drop of $T_c$ occurs at the strain $\eps_{xx}\sim -0.6\%$, while the estimated $\eps_L$ ranges from $-0.75\%$ to $-1.2\%$ (depending on how the first principle calculation is implemented) \cite{scaffidi}. In Ref. \cite{steppke}, the peak in $T_c$ and its sharp drop as the strain pressure further increases are interpreted as arising from the Lifshitz transition in the Fermi surface topology. Our calculations suggest an alternative scenario in terms of the phase transition from SC to SDW$_1$ state.

Finally, we remark that our model for the anisotropic strain is a good approximation to the experiment at small planar strain that conserves the volume. However, experimentally it is possible that $|\eps_{xx}|\neq |\eps_{yy}|$ and $\eps_{zz}\neq 0$ \cite{steppke}. Such experimental details may play a role but make the comparison to theory more complicated and beyond the scope of this work.

To conclude, the recent experiments on the effect of anisotropic strain on the $p$-wave superconductivity in Sr$_2$RuO$_4$ have stimulated us to explore such effects within the single 2D-band picture using SM-FRG. The results are summarized schematically in Fig.1. The initial rise of $T_c$ with strain is consistent with the observation of Steppke {\it et al} \cite{steppke}. Within the SC phase, there is a second transition from T-invariant /T-breaking $p$-wave above/below $T_{c2}$, but the ratio $T_{c2}/T_c$ decreases rapidly with strain, leaving a larger temperature range in the T-invariant phase. The FRG calculation also finds a transition into a small-$\v q$ SDW$_1$ state at higher strains, before the Lifshitz transition is reached, in agreement with the sudden drop of $T_c$ in the experiment. At even larger anisotropic strains the FRG calculation finds a further phase transition into the SDW$_2$ with a larger wavevector $(\pm 1, \pm 0.4)$. Note, in addition to the small wavevectors, neutron scattering also shows strong spin fluctuations at large wavevectors $(\pm 2/3,\pm 2/3)$.\cite{neutron} The latter arise from the 1D $\alpha$ and $\beta$ bands\cite{neutron,others,sro3orb}, which are ignored here since they barely change across $T_c$ and hence are essentially decoupled from superconductivity. Nonetheless our prediction on SDW$_2$ could be modified by interference between these two large wavevector SDW's.  While neutron scattering experiments are likely difficult to perform on the small samples, the prediction hopefully can be tested by NMR experiments.\\

{\em Methods}: The main results presented in Figs.4-6 were obtained numerically by solving the SMFRG equations and the FRG-based mean field theory. Technical details are provided in the Supplementary Materials. \\

{\em Acknowledgments}: We thank Manfred Sigrist for many insightful discussions, Andrew Mackenzie and Clifford Hicks for sharing their experimental data prior to the publication and also stimulating discussions on their experimental results.  We also wish to thank Thomas Scaffidi, Steve Simon and Eun-Ah Kim for interesting discussions and sharing their theoretical works on the strain effect to SC Sr$_2$RuO$_4$. Steppke {\it et al} \cite{steppke} and Scaffidi {\it et al}  \cite{scaffidi} studied the effect of anisotropic strain by one-loop RG within a 3-band model, and their results on the increase of SC $T_c$ are similar to ours. Hsu {\it et al} \cite{kim} considered the effect of uniaxial as well as bi-axial strains on SC. QHW thanks Wan-Sheng Wang for critical reading and comments.\\

{\em Competing interests}: The authors claim no financial conflicts.\\

{\em Contributions}:  QHW, FCZ and TMR designed and developed the project, and YCL performed the calculations. QHW wrote the paper, FCZ and TMR improved the discussions and writing. YCL joined in revising the paper.\\

{\em Funding}: QHW is supported by NSFC (under grant No.11574134) and the Ministry of Science and Technology of China (under grant No. 2016YFA0300401), FCZ is supported in part by National Basic Research Program of China (under grant No.2014CB921203) and NSFC (under grant No.11274269).


\begin{references}
\bibitem{density} Here we ignore the change of chemical potential with strain. We performed separate calculations by tuning the chemical potential so that the electron density is fixed. The qualitative conclusions are not changed.
\end{references}
\end{document}